\documentclass[prb,twocolumn,showpacsamsmath,amssymb,superscriptaddress]{revtex4}
\usepackage{amsmath}
\usepackage{amssymb}
\usepackage{graphicx}
\usepackage{color}

\newcommand{\be}{\begin{equation}}
\newcommand{\ee}{\end{equation}}
\newcommand{\br}{{\bf r}}
\newcommand{\vf}{\varphi}
\newcommand{\lf}{\left}
\newcommand{\rg}{\right}
\newcommand{\ra}{\rangle}
\newcommand{\la}{\langle}

\newcommand{\bea}{\begin{eqnarray}}
\newcommand{\eea}{\end{eqnarray}}

\begin{document}

\title{Strong parameter renormalization from optimum lattice model orbitals}

\author{Valentina Brosco}
\affiliation{Istituto Officina dei Materiali (IOM) and Scuola Internazionale Superiore di
Studi Avanzati (SISSA), Via Bonomea 265, 34136 Trieste, Italy}
\affiliation{ISC-CNR and Dipartimento di Fisica, Universit\`a di Roma
  ``La Sapienza'', Piazzale Aldo Moro 5, I-00185 Roma, Italy}

\author{Zu-Jian Ying}
\affiliation{ISC-CNR and
Dipartimento di Fisica, Universit\`a di Roma ``La Sapienza'', Piazzale Aldo
Moro 5, I-00185 Roma, Italy}
\affiliation{Beijing Computational Science Research Center, Beijing 100084, China}
\affiliation{CNR-SPIN and Dipartimento di Fisica ``E. R. Caianiello'', Universit\`a di Salerno, I-84084 Fisciano (Salerno), Italy}

\author{Jos\'e Lorenzana}
\affiliation{ISC-CNR and
Dipartimento di Fisica, Universit\`a di Roma ``La Sapienza'', Piazzale Aldo
Moro 5, I-00185 Roma, Italy}

\date{\today}

\begin{abstract}
We revisit the old problem of which is the best single particle basis
to express a Hubbard-like lattice model. A rigorous variational
solution of this problem leads to equations in which the answer
depends in a self-consistent manner on the solution of the lattice
model itself. Contrary to naive expectations, for arbitrary small
interactions, the optimized orbitals differ from the
non-interacting ones, leading also to substantial changes in the model
parameters as shown analytically and in an explicit numerical
solution for a simple double-well one-dimensional case.  At strong coupling,
we obtain the direct exchange interaction with a very large
renormalization with important consequences for the explanation of
ferromagnetism with model hamiltonians.
Moreover, in the case of two atoms and two fermions we show that  the
optimization equations are closely related to reduced density matrix
functional theory thus establishing an unsuspected
correspondence between continuum and lattice approaches.
\end{abstract}

\maketitle
\section{Introduction}

The question of what is the most appropriate basis to describe the
electronic Hilbert space of a solid dates back to the original work of Slater and Schockley\cite{slater1936}.
There the authors provide a unified description of optical absorption
in ionic insulators reconciling two apparently competing theories
based respectively on  localized excitations \cite{frenkel1931} and on
Bloch bands  absorption. This problem was later considered by
Wannier\cite{wannier1937} who showed the formal correspondence
between extended Bloch functions  and localized ``Wannier
wave-functions''  paving the way to lattice Hamiltonians,
such as the Hubbard model\cite{Hubbard1963,Kanamori1963,gutzwiller1963} and its
extensions, that are essential to understand many-body phenomena as
Mott insulating behavior, band narrowing,  unconventional
superconductivity and, last but not least, ferromagnetism
which motivated the Hubbard model itself.\cite{gutzwiller1963,Hubbard1963,Kanamori1963} 


The problem of finding the ``best basis'' is very general and
it appears in many different frameworks involving 
electrons,\cite{boys1960,foster1960,edmiston1963,marzari2012,chiesa2013}
 ultra-cold atoms in optical
lattices\cite{Greiner2002,lewenstein2007} and light-matter interactions.\cite{Ying2015b,Liu2015}
The general idea of lattice Hamiltonians in fermionic systems
is that only a subset of bands
close to the Fermi level are important to describe the relevant many-body
physics. Therefore, one can perform a basis truncation 
to the relevant subset of bands and solve in this reduced subspace.
One problem with this plan is that the Wannier orbitals that define
the model are not unique, with different possible choices of  orbitals
related by unitary transformations.\cite{boys1960,foster1960,edmiston1963,marzari2012} This would not
be an issue if the full interaction matrix were retained in the lattice
model. However another common approximation
is Hamiltonian truncation (HT) {\sl i.e.} to truncate the
electron-electron interaction (and sometimes hopping matrix elements) to short
distances which makes the approximate Hamiltonian basis dependent. \cite{paul2006} Several
criteria have been proposed to choose Wannier orbitals that minimize in
some form the error incurred by HT, like
maximally localized Wannier
orbitals,\cite{marzari2012} analogous  to ``Foster-Boys orbitals'' \cite{foster1960,boys1960,edmiston1963} in
quantum chemistry and Wannier orbitals that minimize the intersite
part of the interaction.\cite{paul2006}

Going one step backwards, all these approaches  assume an initial band
or set of bands and the associated Bloch orbitals as a starting
point. Such set of bands and orbitals are usually taken to be the
output of a previous Kohn-Sham\cite{kohn1965}
Density Functional Theory (DFT) computation with some model functional, like
the Local Density Approximation (LDA). Thus, the bands and
initial Bloch orbitals are the solution of an effective non-interacting problem
that approximates the ground state energy and density. However,
it is not clear whether this is  the most appropriate  set of orbitals
to start with.
In this work we address this problem and we show that these choices are not optimal even in weakly interacting
systems. Indeed, we show both numerically and analytically that optimum lattice model
orbitals (OLMO) are surprisingly different from non-interacting orbitals even
for infinitesimal interactions.  As in Refs.~\onlinecite{Spaek2000,kurzyk2008,kadzielawa2013,ying2015a},
 we start by  deriving  a set of self-consistent equations for the  OLMO. In contrast to previous
 works, however,  we focus on the properties of the \emph{exact} solution of the OLMO equations \emph{before} HT studying their structure both analytically and numerically
 in simple cases.
  Not surprisingly, previous approximate
 solutions were shown to be strongly dependent
 on the HT range. \cite{Spaek2000,kurzyk2008,kadzielawa2013}
 Indeed, it is easy to show that basis optimization after HT may lead
 to nonphysical results as, for example, a tendency to minimize the
 on-site Coulomb interaction. This can be contrasted to the tendency to maximize the  Coulomb interaction when the Wannier
orbitals are optimized following the prescriptions of
Refs.~\cite{marzari2012,paul2006} after HT but in a fixed single
particle subspace. 
 To avoid these problems, in our analysis we consider the full Coulomb operator.
Of course, truncation is allowed when it does not affect significantly
the final result, a condition that can be checked a posteriori
changing the truncation range.


A compelling motivation to study OLMO is that, in the last years,
lattice based approaches have been combined with continuum
approaches  to obtain quantitative predictions.
Such hybrid methods include LDA+U\cite{anisimov1997},
DFT + dynamical mean field
theory\cite{anisimov1997a,lichtenstein1998,kotliar2006}, {\em
  ab-initio} Gutzwiller\cite{Julien2005,Wang2008,Ho2008,deng2009,Schickling2014}.
 These methods point at an accurate quantitative
description of many-body phenomena, in contrast to the mainly
qualitative nature that lattice models had up to some years ago. This
calls for an assessment and a better understanding of the errors
incurred by a non-optimal choice of the orbitals used in
practical computations.

Another motivation is that the OLMO may depend on the number
of fermions leading to interesting physics including
superconductivity\cite{Hirsch1989}. Indeed it turns out that the
optimum basis depends self-consistently on the solution
of the lattice Hamiltonian\cite{Spaek2000}. We will show that in
certain cases neglecting the optimization may lead to grossly
non-physical results. In particular such optimization appears
fundamental to avoid a spurious stabilization of ferromagnetism.

In general, the single-particle basis not only determines the
structure of the lattice model and its accuracy, but more generally
it allows to translate  the information gained on the lattice to the
continuum and {\sl viceversa}. 
Recently we have found\cite{ying2015a} a necessary condition to 
describe Mott behavior or stretched molecular bonds
in DFT and show that it is violated by conventional functionals. 
Furthermore, we have shown how the conventional functional 
results can be amended with lattice model results. 
The procedure involves precisely the OLMO  equations which is 
another motivation to study them.

In this work we discuss the general form of the OLMO equations for a
translationally invariant system. The equations are a formidable
problem since they depend on the exact solution of the lattice model which is very hard
 in itself. In order to make progress and estimate the effect
of using OLMO we restrict to a simplified two-site problem with a contact
potential for the fermions. Such potential is particularly appropriate
for fermionic atoms in optical lattices and serves as a proxy for
the OLMO equations for electrons in a molecule or a solid.

One approach that has emerged in last years as a suitable method to go
beyond DFT methods to study correlated systems is Reduced Density
Matrix Functional Theory\cite{helbig2007,buijse2002,gritsenko2005,lathiotakis2009,goedecker1998,sharma2008,sharma2013}(RDMFT).
 Differently from DFT,  RDMFT takes
the full one-body density matrix as the fundamental variational
object \cite{gilbert1975}. A number of different functionals have
  been proposed and tailored to study molecular\cite{buijse2002,gritsenko2005,lathiotakis2009} or solid state systems\cite{goedecker1998,sharma2008,sharma2013}.
Remarkably, we find a close connection between the OLMO equations,
using the Gutzwiller method as a lattice solver, and RDMFT,  thus
establishing a link
that may foster advances on both sides.

\section{Lattice and continuum models}
\label{sec:generalized-hubbard}

We start from a model of fermions in a periodic potential that may be
applicable to electrons in solids or cold atoms in optical lattices,
\begin{eqnarray}
&  H_C&=\sum_\sigma\int \Psi^\dag_\sigma(\br)\hat
  h(\br)\Psi_\sigma(\br)d\br \\
&+&\frac\lambda2\sum_{\sigma,\sigma'}\int\!\int\! \Psi^\dag_\sigma(\br) \Psi^\dag_{\sigma'}(\br')w(\br-\br')\Psi_{\sigma'}(\br') \Psi_\sigma(\br) d\br d\br' .\label{eq:hcont}
 \nonumber
\end{eqnarray}
In the above equation the fields $\Psi_\sigma(\br)$ and $\Psi_\sigma^\dag(\br)$ respectively create  and destroy fermions at position $\br$ with spin $\sigma$,
 $\hat h(\br)$ is the standard one-body Hamiltonian, i.e. $\hat h(\br)=
 -\frac12 \nabla^2+V_{\rm ext}(\br)$, $V_{\rm ext}$ being an external
  static potential,  $w(\br-\br')$ denotes the interaction
  potential among fermions.  $\lambda $ is a dimensionless  coupling
  constant introduced for bookkeeping of the order of the interaction
  in a perturbative expansion and to be set to one at the end of the computation.

To derive the  lattice model we
expand, as usual,  the field operator in an orthogonal
 single particle
basis formed by the set $\{\varphi_{i\sigma},\chi_{\nu\sigma}\}$
where $\varphi_{i\sigma}$ are a set of Wannier orbitals that will be optimized and $\chi_{\nu\sigma}$ are the rest of the states that make the single
particle basis complete. For simplicity we assume one orbital per site
labeled by $i$. The fields are then expressed as,
$$\Psi_\sigma(\br)=\sum_{i} c_{i\sigma}\varphi_{i\sigma}({\bf
  r})+\sum_{\nu}\tilde c_{\nu\sigma}\chi_{\nu\sigma}({\bf r}).$$
with  $c_{i\sigma}$, $\tilde c_{\nu\sigma}$ denoting the fermion destruction operators in
the respective orbitals.

The approximation to a lattice model representation is obtained
by truncating the basis to only the $\varphi_{i\sigma}({\bf
  r})$ states. This  yields the following lattice Hamiltonian,
\begin{equation}
  \label{eq:latmod}
H=\sum_{ij\sigma}h_{ij}
 c_{i\sigma }^{\dag}c_{j\sigma} +\frac\lambda2\sum_{ijkl\sigma\sigma'}w_{ij,kl}
c_{i\sigma}^{\dag}c_{k\sigma'}^{\dag}c_{l\sigma'}c_{j\sigma}
\end{equation}
where we set $h_{ij}=\la\varphi_i|\hat h|\varphi_j\ra$ and the interaction integrals are defined as,
\begin{equation}
  \label{eq:coulint}
w_{ij,kl}   =   \int d^3{\bf r} d^3{\bf r}'\varphi_i^*({\bf r})\varphi_k^*({\bf r}') w({\bf r}, {\bf r}')
\varphi_l({\bf r}')\varphi_j({\bf r}).
\end{equation}
As mentioned in the introduction, we truncate the basis but we
do not truncate the resulting Hamiltonian
operator as it is usually done to simplify the lattice model.\cite{Hubbard1963,Kanamori1963,gutzwiller1963}  Therefore
unitary transformations among the $\varphi_{i\sigma}({\bf
  r})$ orbitals\cite{paul2006} are irrelevant for us as they
corresponds to different representations of the same Hamiltonian
operator.
On the other hand, the choice of the set {$\varphi_{i\sigma}({\bf
  r})$} or, more precisely, the determination of the single particle
Hilbert subspace spanned by {$\varphi_{i\sigma}({\bf
  r})$} is, instead, crucial to obtain an accurate lattice Hamiltonian and it is our central problem.

\section{Optimized basis sets for general lattice models}
\label{sec:opt-gut}

We optimize the orbitals variationally. The energy depends on the lattice wave-function, $|\Phi_S\ra$, and on the single-particle basis states, \{$\varphi_i$\} and it reads
\begin{eqnarray}
  \label{eq:varener}
E[\varphi_i,\varphi_i^*,\Phi_S]&=&E_{\rm 1b}[\varphi_i,\varphi_i^*,\Phi_S]+W[\varphi_i,\varphi_i^*,\Phi_S]\nonumber\\
&&+\sum_{ij}\Omega_{ij}(\la\varphi_i|\varphi_j\ra-\delta_{ij})
\end{eqnarray}
where  $\Omega_{ij}$ is  an Hermitian matrix of Lagrange parameters that
implements the constraint of the orthonormality of the orbitals while the 1-body and interaction contributions to the energy are given by
\bea
\!\!\!\!\!\!\!\!\!\!& & E_{\rm 1b}[\varphi_i,\varphi_i^*,\Phi_S]=\sum_{ij\sigma}h_{ij}\rho_{ij} \label{eq:T+V}\\
\!\!\!\!\!\!\!\!\!\!&
&W[\varphi_i,\varphi_i^*,\Phi_S]=\frac\lambda2\sum_{ijkl\sigma\sigma'}w_{ij,kl}D_{ij,kl} \label{eq:W}
\eea
where $w_{ij,kl}$ and $h_{ij}$ were defined in Eq \eqref{eq:coulint}
and we introduced the spin unresolved one- and two-body density matrices given
respectively by
\bea
 \rho_{ij}=\sum_{\sigma}\la c_{i\sigma }^{\dag}c_{j\sigma} \ra
 &=&\sum_{\sigma}\la \Phi_S|\,c_{i\sigma }^{\dag}c_{j\sigma}\,
 |\Phi_S\ra,\\
D_{ij,kl}=\sum_{\sigma\sigma'}\la c_{i\sigma }^{\dag}c_{k\sigma'
}^{\dag}c_{l\sigma'}c_{j\sigma} \ra &=&\sum_{\sigma\sigma'}\la
\Phi_S|c_{i\sigma }^{\dag}c_{k\sigma'
}^{\dag}c_{l\sigma'}c_{j\sigma} |\Phi_S\ra.\nonumber
\eea

We can perform the minimization of the functional in two steps. First
we minimize with respect to $|\Phi_S\ra$ keeping the orbitals $\varphi_i$ fixed. Then
Eq.~\eqref{eq:varener} yields the expectation value of the lattice
Hamiltonian Eq.~\eqref{eq:latmod} with fixed Hamiltonian matrix
elements (depending on $\varphi_i$). By Ritz variational principle
the minimum is given by the ground state of the lattice Hamiltonian,
\be\label{eq:latground}
H|\Phi_S\ra=E_0|\Phi_S\ra
\ee
where $E_0=E_{1b}+W$.
%

In a second step one performs the minimization with respect to the
wavefunctions $\{\varphi_i\}$ at fixed $|\Phi_S\ra$  and one obtains the following equations
\be
\frac{\delta E[\varphi_i,\varphi_i^*,\Phi_S] }{\delta \varphi_i^*(\bf r)}=0 \quad \frac{\delta E[\varphi_i,\varphi_i^*,\Phi_S] }{\delta \varphi_i(\bf r)}=0.\label{eq:varEphi}
\ee
Using the explicit expression of $E_{\rm 1b}$ and $W$ given above, the
equations can be rewritten in the form
\be \label{eq:varEphi-exp}
\sum_j
\lf(\hat h(\br)\, \rho_{ij} +\lambda \sum_{kl}w_{kl}({\bf r})D_{ij,kl}-\Omega_{ij}\rg)\varphi_j({\bf r})=0.
\ee
Here we introduced the potentials,
$$
 w_{kl}({\bf r})=\int d^3{\bf r'} \varphi_k^*({\bf r}') w({\bf r},{\bf r}') \varphi_l({\bf r}').
$$
The searched minimum is obtained by solving Eqs.
\eqref{eq:latground} and  \eqref{eq:varEphi-exp}
simultaneously. Thus the solution of the many-body problem
depends on the orbitals and {\em viceversa}.

The equations  can be
solved iteratively starting from an initial guess for the lattice
ground states and the wave-functions $\{\varphi_i\}$.  Specifically,
for a given lattice ground-state, Eqs.~(\ref{eq:varEphi-exp}),
represent a set of closed integro-differential equations that can be solved numerically.

One can make a unitary transformation among the orbitals to natural
orbitals defined by the condition that $\rho$ is diagonal.
We will denote matrix elements in the natural orbital basis with
a bar, thus  $\bar\rho_{\mu\nu}=\delta_{\mu\nu}\bar\rho_\mu$.
In the case of a translational invariant system, the natural orbitals
are Bloch orbitals and by symmetry also $\Omega$ becomes
diagonal yielding simpler expressions,
\be
 \label{eq:varEphi-natorb}
\lf(\hat h(\br)\, \bar \rho_{{\bf k}}- \bar\Omega_{{\bf k}}\rg)\!\psi_{\bf k}({\bf r})+\lambda\sum_{{\bf k}' {\bf q}} \bar D_{{\bf k}{\bf k}'{\bf q}} \bar w_{{\bf k}' {\bf k}+{\bf q}}({\bf r})\psi_{{\bf k}'-{\bf q}}({\bf r})=0.
\ee
In the above equation  $\psi_{\bf k}({\bf r})$ indicates  the Bloch
state with momentum $\boldsymbol{{\bf k}}$  defined as
$$\psi_{\bf k}({\bf r})=\frac{1}{\sqrt{L}}\sum_j \varphi_j({\bf r})
e^{-i\,\boldsymbol{{\bf k}} {\bf R}_j},$$  where $L$ denotes the number of
lattice sites and ${\bf R}_j$ indicates the position of the $j$-th
site,  $\bar w_{{\bf k},{\bf k}'}({\bf r})$,  $\bar\Omega_{{\bf k}}$
$\bar D_{{\bf k},{\bf k}',{\bf q}}$ and $\bar \rho_{{\bf k}}$ are the
corresponding quantities in the basis of Bloch
states.

%

\section{Two-site problem}
For a two-site potential with inversion symmetry (a homoatomic
molecule in the electronic case) the minimal single-particle basis
consists of two states. It is again convenient to go to a natural
orbital basis where both  $\rho$ and $\Omega$ are diagonal. There are
two natural orbitals which can be classified by parity
 $\psi_0$ (even) and $\psi_1$ (odd).
We can define  Wannier orbitals,
\begin{eqnarray}
  \label{eq:wa2}
 \vf_a&=&\cos(\theta)  \psi_0 + \sin(\theta)  \psi_1\nonumber\\
 \vf_b&=&\sin(\theta) \psi_0 - \cos(\theta) \psi_1.
\end{eqnarray}
By requiring that the Wannier orbitals preserve the symmetry of the
problem $\vf_a(\br)=\vf_b(-\br)$ one gets $\theta=\pi/2$ so in this
case the Wannier orbitals are uniquely determined by the natural orbitals.

The lattice Hamiltonian can be easily diagonalised, the spectrum
consists of six states, a degenerate triplet corresponding to $S=1$
and $S_z=1,-1,0$ and three singlets  corresponding to different
combinations of the Heitler-London state and  of the two ionic
states. The states corresponding to the  two lowest energy eigenvalues are a singlet and the degenerate triplet.

\subsection{Optimization equations for a singlet ground-state}
The lowest energy singlet  of the model Hamiltonian Eq.~\eqref{eq:latmod} has the form
\begin{equation}
  \label{eq:gwfa}
|\Phi_S
\rangle=\frac{1}{C_\gamma^{1/2}}\left[(\gamma+1)a_{0\uparrow}^\dagger
a_{0\downarrow}^\dagger
+(\gamma-1) a_{1\uparrow}^\dagger
a_{1\downarrow}^\dagger \right]|\emptyset\rangle,
\end{equation}
where $a_{0\sigma}$ and $a_{1\sigma}$ destroy a fermion  with spin $\sigma$ in the orbital $\psi_0$ and $\psi_1$ and $C_\gamma= (1+\gamma^2)$.

$\Phi_S$ has the form  of
variational Gutzwiller wave-function\cite{gutzwiller1963,gebhard2009}  and for a wide range of parameters is the exact lattice ground-state.
$\gamma$ denotes the Gutzwiller variational parameter  and it allows a smooth
interpolation between the Hartree-Fock (HF) and the Heitler-London (HL)
wave-functions ($\gamma=1$ and $\gamma\rightarrow 0$ respectively).
The optimum value of $\gamma$ for general matrix elements is given by
\begin{equation}
\gamma_{\rm opt} =\frac{\Delta _{U}-\lambda(U-V-K+K')}{4(t-\lambda t_c) }
\label{eq:gamma}
\end{equation}
with $\Delta_{U}=\sqrt{\lambda^2\left( U-V-K+K^{\prime }\right) ^2+16\left(
t-\lambda t_c\right) ^2}$ where $t=h_{ab}$
%
while $U, V,\,K,\,K^{\prime }$ and $t_c$ are given by the following equations in the basis spanned by the orbitals $\varphi_a$
 and $\varphi_b$
\begin{eqnarray}
  \label{eq:genhub}
& &U= w_{aa,aa}, \quad  V= w_{aa,bb}, \quad  t_c= w_{aa,ab},\nonumber\\
& & K= w_{ab,ba},\quad K'=  w_{ab,ab}.\label{eq:k}
\end{eqnarray}
 By looking at the above equations we see that $V$ denotes the inter-site repulsion, $K$ is the direct
exchange interaction, $K'$ can be thought as a
repulsion among bond-charges and $t_c$ can be considered as the contribution of the Hartree potential to the hopping.
For real orbitals we have   $K=K'$ and, in the basis $\psi_0, \psi_1$
we can express $\gamma_{\rm opt}$ simply as a function of $U-V=2\bar
w_{01,01}$, $t$ and $t_c=(\bar w_{00,00}-\bar w_{11,11})/4$, where
$\bar w_{\lambda\mu,\nu\rho}$,   indicates the matrix elements of the interaction in the natural orbital basis.

Using Eqs.(\ref{eq:gwfa}) it is straightforward to obtain the full explicit form of the energy  functional in the two-site case:
\bea E_{\rm
  2site}[\psi_0,\psi_1,\gamma]&=&\sum_{\mu}\bar\rho_{\mu}\big(\bar h_{\mu\mu}+\frac{\lambda}{2}\bar w_{\mu\mu,\mu\mu}\big)+ \nonumber\\
& &\!\!\!\!\!\!\!\! -\lambda\frac{\sqrt{1-q^2}}{2}(\bar w_{10,01}+\bar w_{01,10}). \label{varGutz} \eea
with $\bar\rho_0=1+q$, $\bar\rho_1=1-q$ and $q=2\gamma/(1+\gamma^2)$.

Starting from this equation, applying the optimization scheme
explained in Sec. \ref{sec:opt-gut}, we arrive at the analogous of
equation \eqref{eq:varEphi-natorb} for a two-site system 
\be\label{eq:mineq}
\sum_{\nu} \hat {\cal H}_{\mu\nu}({\bf r})\psi_{\nu}({\bf r})=\Omega_{\mu}\psi_{\mu}({\bf r}) \quad {\rm with}\,\, \mu,\nu\in[0,1]
\ee
where
\be
\label{eq:calH}
\hat {\cal H}({\bf r})=\lf( \begin{array}{cc}
   [\hat h({\bf r}) +\lambda\, \bar w_{00}({\bf r})]\bar\rho_0 & - \lambda\,\bar w_{01}({\bf r})\sqrt{\bar\rho_0\bar\rho_1} \\
   - \lambda\,\bar w_{10}({\bf r})\sqrt{\bar\rho_0\bar\rho_1} &  (\hat h({\bf r}) +\lambda\,\bar w_{11}({\bf r}))\bar\rho_1 \\
  \end{array}\rg).\ee
Incidentally we  notice that this matrix has a very simple structure,
\be \hat {\cal H}=\sqrt{\rho}\,(\hat h(\br)\, {\cal I}+\lambda\,{\cal W}({\bf r}))\,\sqrt{ \rho}\label{eq:calH2s}\ee
where the square root is intended in the operator sense, {\sl i.e.} in the natural orbital basis $\sqrt{\bar\rho}_{\mu\nu}=\sqrt{\bar\rho_\mu}\delta_{\mu\nu}$, ${\cal I}$ denotes the $2\times2$ identity matrix and the matrix ${\cal W}$ is defined as
\be\lf[{\cal W}(\br)\rg]_{\mu\nu}=f_\mu f_\nu \bar w_{\nu\mu}(\br)\label{eq:calW2s}\ee
with  $\bar w_{\mu\nu}(\br')=\int d\br\, \psi_{\mu}^*(\br)\, w(\br-\br')\,\psi_{\nu}(\br)$ and $f_0=1$ and $f_1=-1$.
From Eqs. \eqref{eq:calH2s} and  \eqref{eq:calW2s} we can recover
Gutzwiller variational energy by doing
\be E_G=\sum_{\mu}\bar\rho_{\mu}\bar h_{\mu\mu}+\frac\lambda2\sum_{\mu\nu}\la \psi_\mu | \lf[\sqrt{\rho}\,{\cal W}\sqrt{\rho}\rg]_{\mu\nu}|\psi_\nu\ra\label{gutzwiller-energy}.\ee

A nice additional feature of these equations is that the eigenvalues
$\bar\Omega_\mu$ can be put in an approximate relation with the ionization
energy, $I$. We indeed find that the following relation holds:
$\bar\Omega_\mu=\bar\rho_\mu(E_G-\bar h_{\mu\mu})$ which implies that,
if we  neglect the relaxation of the orbitals setting $\bar
h_{00}\simeq E_{\rm 1fermion}$,  we have $\bar\Omega_0=-\bar\rho_0 I$.

To understand their physical meaning, it is interesting to look at the optimization equations in different limits.

\subsubsection{Limit of weak interaction}\label{weak-int}

We first analyze the limit of weak-interaction realized for
$\lambda\rightarrow0$ and finite $R$.
As we now show, the equations for a vanishingly small interaction
differ from the non-interacting Schr\"odinger equation for the
orbitals by a finite amount. 

At zero interaction, we have ${\cal
  W}=0$, $\bar\rho_0=2$ and $\bar\rho_1=0$. The equation  for
$\psi_1$ [Eq.~\eqref{eq:mineq}] is therefore trivially satisfied by choosing $\bar\Omega_1=0$ and  $\psi_1$  is undetermined;  while the equation for $\psi_0$ simply reduces to the equation for the non-interacting ground state, {\sl i.e.} $\psi_0$ coincides with  the bonding molecular eigenstate while $\bar\Omega_0=2 \varepsilon_g$.

To properly describe the limit of non-zero but very weak-interaction,
we expand the 
matrix ${\cal H}$ to lowest  order in the small parameter
$\lambda$. Noticing that $\rho_0=2+O(\lambda^2)$,
$\rho_1=O(\lambda^2)$ and setting $\bar\Omega_\mu=\bar\rho_\mu\epsilon_\mu$, we obtain the following equations valid to zeroth order in $\lambda$:
\bea
& &\!\!\!\!\!\!\!\!\!\!\!\!\!\! \hat h(\br)\, \psi_0(\br) =\epsilon_0\psi_0(\br) \label{eq:wi-psi0}\\
& & \!\!\!\!\!\!\!\!\!\!\!\!\!\!  \hat h(\br)\,\psi_1(\br)- \frac{8 t}{U-V}\,\bar w_{01}(\br)\psi_0(\br)=\epsilon_1\psi_1(\br).\label{eq:wi-psi1}
\eea

We see that the equation for $\psi_0$
yields, as expected, the bonding molecular eigenstate while the
equation for  $\psi_1$ can be seen as a Schr\"odinger equation 
with a non-local potential depending on  $\psi_0$ (notice that
$\psi_1$ appears in the integral kernel of $\bar w_{01}$). 


More importantly, the non-local potential remains finite for an
infinitesimal interaction ($\lambda\rightarrow 0 $). It follows that, in the limit of vanishing
small interactions; the antibonding state, the optimized Wannier orbitals Eq.~\eqref{eq:wa2} and the
lattice Hamiltonian parameters do not converge to their
non-interacting values.
It is easy to see that this will persist for an extended system and
for arbitrary well-behaved interaction potentials.

This result is easy to understand by
considering the total energy of the system in the non-interacting and in
the interacting case.
 In the non-interacting case the ground-state can be constructed by diagonalizing the single particle Hamiltonian and
occupying the lowest $N_e$ states with $N_e$ denoting the number of
electrons. Now suppose we want to construct a lattice model with $N_o>N_e$ orbitals (counting spin).  Since the energy does not depend on
unoccupied orbitals any basis of Wannier orbitals 
constructed from the $N_e$ occupied orbitals and any set of $N_o-N_e$
unoccupied orbitals is acceptable. Indeed, any such choice leads to 
a parametrization of the  Hamiltonian that reproduces the non-interacting ground state
energy. Thus, in the absence of interaction, the OLMO procedure does
not lead to a unique  Wannier orbitals basis.

The situation changes for an infinitesimal interaction because now all
 orbitals have a finite occupation and they determine the ground state energy. In this case the OLMO
procedure  picks out a well-defined set of orbitals that in general
differ from  the non-interacting ones.
%
%
%


\subsubsection{Limit of large distances}

Another interesting case is the limit of  large distances  where
$\frac{\lambda U}{4 t}\gg 1$ and the singlet ground-state
approximately coincides with the HL ground-state.
In this case, to lowest order, $\bar\rho_0=\bar\rho_1=1$ and the
optimization equations for the orbitals read,
\begin{eqnarray}
\hat h(\br) \psi_0+ \lambda [\bar w_{00}(\br)\psi_0-\bar w_{01}(\br)\psi_1]&=&\bar\Omega_0\psi_0,\label{eq:si-psi0}\\
\hat h (\br)\psi_1+  \lambda [\bar w_{11}(\br)\psi_1-\bar w_{01}(\br)\psi_0]&=&\bar\Omega_1\psi_1.\label{eq:si-psi1}
\end{eqnarray}
Differently from the weak-interaction limit, we have two coupled
non-local equations for  $\psi_0$ and $\psi_1$ which can be solved numerically.
We have used these equations to check our general numerical solution below.

\subsection{Optimization equations for a triplet  ground-state}
The energy of the triplet can be written as
\be E_{\rm triplet}= h_{00} + h_{11}+\lambda(V-K).\ee 

Starting from this expression it can be easily shown that the OLMO
satisfy the following optimisation  equations
\begin{eqnarray}
\hat h(\br) \psi_0+ 4\lambda [\bar w_{11}(\br)\psi_0-\bar w_{01}(\br)\psi_1]&=&\epsilon_0\psi_0,\label{eq:t-psi0}\\
\hat h(\br) \psi_1+ 4\lambda [\bar w_{00}(\br)\psi_1-\bar w_{01}(\br)\psi_0]&=&\epsilon_1\psi_1.\label{eq:t-psi1} 
\end{eqnarray}
To arrive at the above equations  we used  the following relation, stemming from Eqs.\eqref{eq:genhub},
$[V-K]=2\lf(\bar w_{11,00}-\bar w_{01,01}\rg)$.

\section{Numerical solution for a simple two-atom molecule}
%

We now explicitly calculate  the optimized Wannier orbitals  for a
one-dimensional system of two fermions interacting {\sl via} a repulsive contact interaction in an external two-well potential. A system of this kind has been recently realized with ultracold atoms \cite{murmann2015} and it may be used as a proxy
to have a qualitative understanding of the OLMO for electrons. 

For the single-particle Hamiltonian we use a  model  first introduced by A. Caticha \cite{caticha1995}. This model has two nice features: first the bonding and antibonding eigenstates are exactly known and
they can be calculated by means of a  superpotential, as it is done in supersymmetric quantum mechanics\cite{cooper1995}; second, in the limit of large inter-well distances, it
reduces to a superposition of two Eckart  wells\cite{landau} allowing to recover a model often used in molecular physics\cite{helbig2009}.

\begin{figure}[t]
\begin{flushleft}
\includegraphics[width=0.45\textwidth]{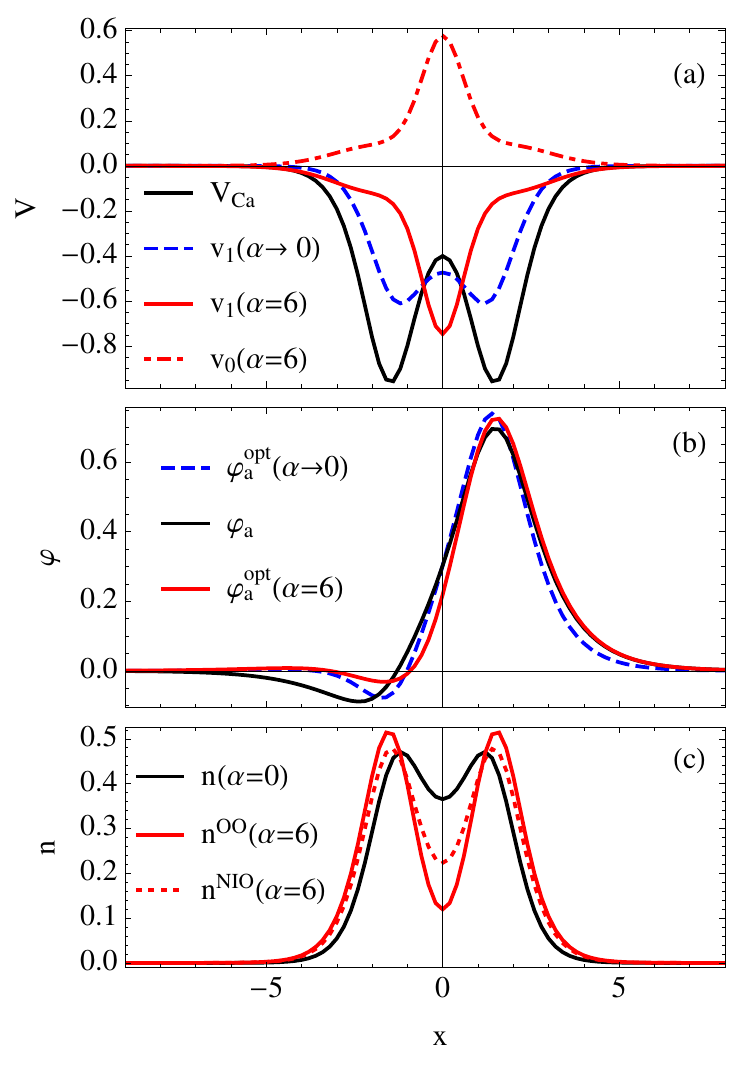}
\caption{(a) Caticha potential, $V_{\rm Ca}(x;a,b)$, along with the effective potentials, $v_0$ and $v_1$, in the limit of vanishing interaction, $\alpha \rightarrow 0$, and for $\alpha=3$. For $\alpha \rightarrow 0$, only $v_1\neq 0$. Other parameters: $t_0=0.2$.
(b)Non-interacting and optimized Wannier states for $\alpha \rightarrow 0$ and $\alpha=3$. (c) Ground state density (red lines) compared  with the non-interacting density (black line).
The solid and dotted lines show the density calculated with optimized and non-interacting orbitals, respectively.}
\label{fig:phi-pot}
\end{flushleft}
\end{figure}

Caticha potential can be constructed starting from two real parameters, below denoted as  $a$ and $b$, that are related to the bonding and antibonding energies, respectively $\varepsilon_g$ and $\varepsilon_u$, as $\varepsilon_g=-a^2$ and $\varepsilon_u=-b^2$.
Measuring energies in units of
$I_0=(a^2+b^2)/2=1$ and distances in units of  $a_0$ with
$a_0^2= \hbar^2/(m I_0)=1$,  the single-particle Hamiltonian then reads,
$\hat h_{\rm Ca}= -\frac{\nabla^2}{2}+V_{\rm Ca}$
where  Caticha   potential  is given by
\be V_{\rm Ca}(x;a,b)=\frac{-t_0\lf(2 t_0+ b^2 \cosh(2 a x)+a^2 \cosh(2 b x)\rg)}{a \cosh(ax)\cosh(bx)-b \sinh(ax)\sinh(bx)} \label{eq:Vca}\ee
with $t_0=(a^2-b^2)/2$ representing the non-interacting tunnelling energy  and $a>b$.
The difference $a-b$ thus controls the distance between the two wells, that as usual scales as $\log(t_0)$.
The solid black line in Figure \ref{fig:phi-pot}(a) shows Caticha potential for $t_0=0.1$.
%
%

For a contact interaction, the orbital optimization equations for the
singlet state $|\Phi_S\ra$,  Eqs.(\ref{eq:mineq}-\ref{eq:calH}), become local and they reduce to a set of
coupled orbital-dependent  single-particle Schr\"odinger equations with self-consistently defined external potentials. Specifically, setting  $w(x-x')=\alpha\, \delta(x-x')$ and $\lambda=1$, for all
 $\alpha\neq 0$ we can write:
%
\be \label{eq:delta-opt}
\hat h(x) \psi_\mu(x)+ v_\mu (x)\psi_\mu(x)=\bar\epsilon_\mu\psi_\mu(x), \quad {\rm with}\, \mu=0,1
\ee
and
\be\label{v0-v1}
v_0= \alpha\!\lf( \psi_0^2-\sqrt{\frac{\bar\rho_1}{\bar\rho_0}} \psi_1^2 \rg) \,\,{\rm and}\,\,
v_1=  \alpha\!\lf(\psi_1^2-\sqrt{\frac{\bar\rho_0}{\bar\rho_1}}\psi_0^2 \rg).
\ee



We solve Eqs. (\ref{eq:delta-opt}-\ref{v0-v1}) numerically. The
results are also shown in Fig.\ref{fig:phi-pot}(a-c).

 We remind that the above equations are for a singlet ground state. In
the case of a triplet the interaction terms cancel in
Eq.~\eqref{eq:t-psi0},~\eqref{eq:t-psi1}, however since both orbitals
are occupied the variational problem is well defined and its solution  simply yields the bonding and anti-bonding 
eigenstates of the single-particle Hamiltonian.
As a result, in the following,``optimized orbitals'' (OO) always refer to the optimum orbitals for a singlet ground-state, while the bonding and anti-bonding 
eigenstates of the non interacting problem 
are called  ``non-interacting orbitals'' (NIO). As mentioned, for the
present contact interaction, they are
also the optimized orbitals in the triplet subspace. 



\begin{figure}[t]
\begin{flushleft}
\includegraphics[width=0.45\textwidth]{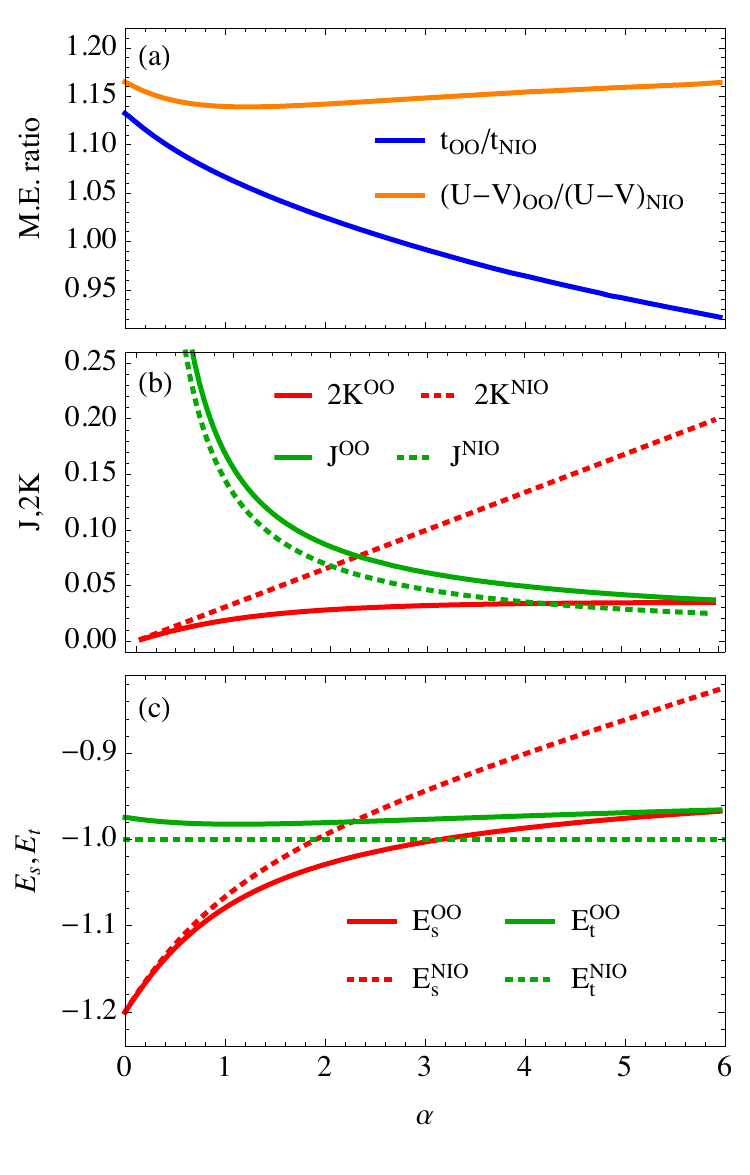}
\caption{
 (a)Renormalization of the hopping amplitude and of $U-V$ as a
 function of $\alpha$ shown as ratio between the matrix element
 (M.E.) computed with optimized orbitals (OO) and non-interacting
 orbitals (NIO). 
 (b) Direct and superexchange couplings as a function of $\alpha$ (c)
 Energy of the singlet and of the triplet  as a function of
 $\alpha$. In both panels (b) and (c) solid and dotted lines indicate
 quantities  calculated using optimized and non-interacting orbitals.
Energies are in units of $I_0$. }
\label{fig:renpar}
\end{flushleft}
\end{figure}

As discussed in Sec.~\ref{weak-int}  even for an infinitesimal
interaction, the optimum $\psi_1$ differs from the non-interacting
antibonding molecular eigenstate. 
It is interesting to obtain an expression for the potential
determining $\psi_1$ in the weak coupling limit of Eq.~\ref{eq:wi-psi1},
$$v_1= -\frac{8t}{1-r_{ab}} \psi_0^2,$$
were $r_{ab}=\int dx \phi_a^2\phi_b^2 $. 
As we see in Fig.~\ref{fig:phi-pot}(a),  in the  presence of an infinitesimal interaction, the potential $v_1$ represented by the dashed blue line, essentially tends to confine the charge closer to the bond region.
The corresponding optimum Wannier orbitals  therefore shrink and shift toward the bond with a consequent sizable increase of the tunneling amplitude and of the Hubbard interaction, $U$ with respect to their values calculated starting from the non-interacting molecular eigenstates, as shown in Fig.~\ref{fig:renpar}(a) where we see that,  even in the vanishing interaction limit, we have a finite renormalisation of $t$ and $U$.

At finite $\alpha$, the optimum orbitals $\psi_0$ and $\psi_1$ are
respectively the ground and first excited state of the single-particle
Hamiltonians defined by Eqs. (\ref{eq:delta-opt}-\ref{v0-v1}). As
shown by the dashed  and solid red lines in Fig. \ref{fig:phi-pot}(a), in this case, 
both  $v_{0}$ and $v_1$ are roughly confined to the bond region and
they have opposite characters, $v_0$ is repulsive while $v_1$ is
attractive.
 $v_0$ becomes important going beyond the weak-coupling limit and its dominant
 effect is that it tends to reduce the weight of  $\psi_0$ in the bond
 region and it leads to a suppression of the tunneling 
 compensating the effect of $v_1$, see Fig.\ref{fig:renpar}(a).


The optimization of  the orbitals not only affects the parameters of
the lattice model but it also leads to different ground state densities when we go back to the continuum, this is what we show in Fig.~\ref{fig:phi-pot}(c). There we plot the density  calculated with optimized and non-interacting orbitals, respectively $n^{\rm OO}=\bar\rho_0^{\rm OO}\psi_0^2+\bar\rho_1^{\rm OO}\psi_1^2$ and $n^{\rm NIO}=\bar\rho_0^{\rm NIO}\psi_g^2+\bar\rho_1^{\rm NIO}\psi_u^2$, along with the  non-interacting density, $n_0=2\psi_g^2$ with  $\psi_g$ and $\psi_u$ denoting the bonding and antibonding non-interacting orbitals and $\bar\rho_\mu^{\rm OO}$ and $\bar\rho_\mu^{\rm NIO}$ indicating respectively the elements of the one-body density matrix calculated with the optimized and with the non-interacting lattice parameters.
We see that there is a substantial difference between the density in the three cases and that the optimization of the orbitals strongly reduces the bond charge. Furthermore by considering Fig.\ref{fig:renpar}(a), one can easily understand that orbital optimization generally tends to increase the ratio $U/t$ thus bringing the system in a more correlated state.

\subsection{Orbital optimization effects for excitation energies and the  phase diagram}

A major purpose of model Hamiltonians is to compute phase diagrams and
excitation energies. For a periodic Mott insulating system one is
interested in the magnetic Hamiltonian. It is customary
to compute the magnetic parameters for a dimer and insert them
 in the many-site
Heisenberg model.\cite{Anderson1950} Proceeding in this way the many-site
Heisenberg model will have a ferromagnetic ground state when the
dimer has a triplet ground state.  Below we will show that the
stability of the triplet state is very sensitive to the strategy used
to obtain the model parameters with important implications for the
description of ferromagnetism with model Hamiltonians.

Once the variational setting of this work is adopted for optimizing
the orbitals the following problem appears. Since one can apply the
variational principle in each subspace with well defined quantum numbers,
if the orbitals depend on
the state, {\em in principle}, one should optimize the orbitals in any 
such subspace, i.e. one should use different orbitals and
parameters depending if one is computing the energy of the ground
state or of an excited state with different quantum numbers (like the
triplet in the present case). Not only this would be very cumbersome,
but we will show below that such strategy leads to unphysical results.
At least in the present case, we can show that, due to a cancellation of
errors,  the best strategy to compute the phase diagram and excitations
energies is to compute the singlet and the triplet energies with the parameters
optimized for the same singlet state.

Let us consider the effects of orbital optimization on the exchange energy defined as usual as the energy difference between triplet and singlet.
For our simple  toy-model the latter  depends on the choice of the orbitals and it is in general given by:  
%
%
\be E_t-E_s=\frac{\Delta_{{\rm U}}-U+V-4K}{2}-\Delta_h\label{EST2}\ee
%
where   $\Delta_h=0$ if the singlet optimized orbitals are used for both the
singlet and the triplet wave-functions and energies.  Hereafter this
choice will be dubbed ``fixed
singlet orbitals''  (FSO).  Another possibility is to use  ``state
dependent orbitals'' (SDO). In this case
parameters for each state are computed with its own optimized orbitals
and $\Delta_h$
accounts for the effects of orbital relaxation, {\sl i.e.}  $\Delta_h=2 I_0+\bar h_{00}+\bar h_{11}$
with the last two terms
computed using the singlet optimized orbitals. 
In both cases, FSO and SDO,  the Coulomb matrix elements appearing in
Eq.~\eqref{EST2} have to be calculated using the singlet-optimized
orbitals since, for a contact interaction, the Coulomb matrix elements do not contribute to the triplet's energy.

Notice that for FSO and assuming $U-V \gg t,\,K,\,t_c$,    we  recover the standard result that the exchange equals the sum of a kinetic and a Coulomb term and it is approximately given by:

%

%
\bea E_t-E_s&\sim& \frac{2(t-t_c)^2}{U-V}-K=J/2-K.\label{EST3}\eea
where we indicated as $J$  the super-exchange interaction,
$J=4(t-t_c)^2/(U-V)$. Thus, for sufficiently large direct exchange the
ground state becomes a triplet.

In Fig. \ref{fig:renpar}b we show the effects of orbital optimization
on  $J$ and $K$ as a function of $\alpha$ 
for the optimized orbitals
assuming the singlet ground state (OO). We also show the result for
non-interacting orbitals (NIO) which coincides with the parameters
that would be obtained for the triplet ground state.

As one can see, while the
changes in $J$ are small or moderate $K$ is strongly renormalized by
orbital optimization in the singlet state. In
particular, while in the absence of orbital optimization $K$ grows
linearly with $\alpha$, in the presence of singlet orbital optimization it
tends to saturate.

A spurious singlet to triplet transition is found for $\alpha\sim 2$
if the parameters are computed with the widely used approximation of
taking non-interacting orbitals.
This is also what we see in Fig. \ref{fig:renpar}c where we plot the energy of the singlet and triplet as a function of $\alpha$.

The transition is shifted to a larger value $\alpha\sim 3$ if one
adopts the SDO strategy and computes the ground state energy (panel c,
full red line for the singlet and dashed green line for the triplet). A much improved result is obtained if the FSO
strategy is applied (full lines in panel c). In this case the transition
disappears from our studied window and singlet and triplet become degenerate at
strong coupling as expected from the exact solution.\cite{girardeau1960}
In this approximation one can use a fixed model Hamiltonian to compute
ground state and excited states as customary done which is obviously a
great simplification.

%


%

In these regards let us remark that in the limit $\alpha\rightarrow\infty$, $|\phi_S\ra$ does not represent the lowest energy singlet, indeed it can be shown by means of the so-called Girardeau mapping\cite{girardeau1960} that the correct singlet ground-state should have the form:
\be |\Phi_{S,\infty}\ra=|\psi_0(\br_1)\psi_1(\br_2)-\psi_1(\br_1)\psi_0(\br_2)|. \ee
%
This state is degenerate with the triplet and it lies outside the space of our trial wave-functions, that is essentially restricted to the all the states that can be expressed as the sum of Slater determinants constructed with the optimized orbitals.
More precisely, the state $|\Phi_{S,\infty}\ra$ has a non-analytic behavior at $\br_2=\br_1$, which could be described only by an infinite sum of Slater determinants and cannot be captured by the our trial wavefunction.
Indeed, our trial wave-function is just the best variational state
within the restricted basis and contains only two Slater
determinants [Eq.~\eqref{eq:gwfa}].  The situation is different for
the triplet state which for the spatial part reads,
\be |\Phi_{T,\alpha}\ra=\psi_0(\br_1)\psi_1(\br_2)-\psi_1(\br_1)\psi_0(\br_2). \ee
This state coincides with our trial triplet state 
and is the exact ground state for all values of $\alpha$ in the
triplet subspace. 
The triplet energy $E_t=-I_0$
appearing in Fig.~\ref{fig:renpar}c is the exact triplet energy and
also the asymptotic ground state energy for
$\alpha\rightarrow\infty$ were the triplet and the singlet become
degenerate.  

All the spurious transitions reported above are rooted in
the fact that the variational state yields the exact ground state
energy for the triplet and a variational upper bound energy for the
singlet. When computing energy differences errors do not cancel and the triplet
is spuriously favored. 
 A similar problem arises in lattice Hamiltonians when trying
to describe ferromagnetism.\cite{Gunther2011}

Using the FSO  strategy
the same systematic error is incurred in the computation of the ground state
energy of the triplet and the singlet so that the error cancels in the
excitation energy and a physically correct value is obtained. Thus, for
example, the spin wave spectrum will be physically meaningful for a
Heisenberg model with parameters computed within the  FSO scheme while it will give a generally
wrong result if orbital relaxation is neglected or if the SDO strategy
is adopted to compute an effective exchange energy. The price to pay
is that a spurious shift appears in the total energy of the system
however this shift is irrelevant to compute most physical properties
with the model Hamiltonian.


\section{Gutzwiller natural orbital functional }
\label{sect:ext-sys}
As we discussed so far, for a  two-site molecule
 the structure of the one- and two-body density matrices, $D$ and $\rho$, appearing in Eqs. (\ref{eq:varEphi-exp}) is exactly known.
 This allowed us to write the  energy as a functional of a single
 parameter $\gamma$,  which determines the lattice ground-state, and
 of  the orbitals $\{\varphi_i\}$, see Eq. \eqref{varGutz}.

In very rough terms, the logic behind Gutzwiller {\sl Ansatz} can be summarized by the following statement:
a lattice system with a repulsive Hubbard interaction should have a
smaller doubly occupancy probability than a non-interacting one.
The variational Gutzwiller wave function for the lattice model can be
written as,\cite{gutzwiller1963,gebhard2009,Vollhardt1984,Metzner1987,Gebhard1990}
\be|\Phi_\gamma \rangle=\frac{\gamma^D}{C^{1/2}_\gamma}|\Phi_0 \rangle \label{eq:psigamma}\ee
where $|\Phi_0 \rangle$ is a Slater determinant, $D=
\sum_i n_{i\uparrow }n_{i\downarrow }$ counts the total double
occupancy, $\gamma$ is a variational parameter and $C_\gamma=\langle
\Phi_0 |\gamma^{2D}|\Phi_0 \rangle$ a normalization constant.
In the two-site case this equation simply gives the exact lattice ground-state of Eq. \eqref{eq:gwfa}.

The projector $\gamma^D$ reduces the weights of the configurations with doubly occupied sites and  the optimization of $\gamma$ essentially coincides with the optimization of the average  double occupancy.

A somewhat complementary
point of view is adopted in  RDMFT
\cite{helbig2007,buijse2002,gritsenko2005,lathiotakis2009,goedecker1998,sharma2008,sharma2013}
where the fundamental variational object is the whole one-body density matrix\cite{gilbert1975} defined by its eigenstates and the corresponding eigenvalues as
\be
\Gamma(\br,\br')=\sum_{n}\phi^*_n(\br)\phi_n(\br')\bar\rho_n.
\ee

Since within Gutzwiller approach $\gamma$ completely determines,
through $q$, the full one-body density matrix on the lattice, one may
``eliminate'' $\gamma$ in favor  of $q$ and  extend Gutzwiller theory
to the continuum by constructing a RDMFT which reduces to standard
Gutzwiller theory when the single particle basis is kept fixed but
which determines the orbitals variationally in the general case.
In the following, we show how the above reasoning  applies to the two-site system.

In the two-site case, starting from the lattice Gutzwiller wave-function Eq. \eqref{eq:gwfa},  it is
straightforward to obtain the corresponding 
one- and  two-body density matrices in
terms of the bonding and antibonding orbitals, $\psi_0$ and $\psi_1$, and of their occupancies. We indeed have
for the one-body density matrix
\begin{eqnarray}
\Gamma_1^{\rm G}(\br,
\br')&=&\sum_{\sigma}\la\Phi_\gamma|\Psi^\dag_{\sigma}(\br)\Psi_\sigma(\br')|\Phi_\gamma\ra\label{gamma1} \\
&=&(1+q)\psi_0(\br)\psi_0(\br')+(1-q)\psi_1(\br)\psi_1(\br')\nonumber
\end{eqnarray}
where $q=2\gamma/C_{\gamma}$
while for the diagonal part of the  two-body density we obtain:
\begin{eqnarray} \label{gamma2}
&\Gamma^{\rm G}_{2}(\br,
\br')=\sum_{\sigma\sigma'}\la\Phi_\gamma|\Psi_{\sigma'}^\dag(\br')\Psi_\sigma^\dag(\br)\Psi_\sigma(\br)\Psi_{\sigma'}(\br')|\Phi_\gamma\ra&\nonumber\\
&=\lf(\sqrt{1+q}\,\psi_0(\br)\psi_0(\br')-\sqrt{1-q}\,\psi_1(\br)\psi_1(\br')\rg)^2.&
\end{eqnarray}
By direct comparison we easily see that, in a ``minimal''  basis consisting of just two orbitals, $\Gamma^{\rm G}_{2}(\br, \br')$ coincides with the exact two-body density for a two-fermion closed-shell system first obtained by L\"owdin and Shull (LS)\cite{lowdin1956}.
Equations \eqref{gamma1} and \eqref{gamma2} therefore  show that for a two-site molecule Gutzwiller theory just gives the  LS density matrix
functional in a restricted basis.
More explicitly by replacing Eqs. (\ref{gamma1}-\ref{gamma2}) in the well-known expression of the energy in terms of the one-  and two- body density matrices, one can easily show that the following relation holds
\bea E[\Gamma^{\rm G}_{1},\Gamma^{\rm G}_{2}] \!&=&\!\!\!\int \!d{\bf
  r'}  \lim_{\bf r \rightarrow \bf r'}\lf[ \hat h_{\bf r}\Gamma^{\rm
  G}_{1}(\br, \br')\rg]+\!\nonumber\\ & \!\!\!\!\!\!\!\!\!
\!\!\!+&\!\!\!\!\!\!\!\!\! \!\!\!\frac{1}{2}\!\int \!\Gamma^{\rm
  G}_{2}(\br, \br') w(\br, \br') d{\bf r} d{\bf r'}=E_{\rm
  2site}[\psi_0,\psi_1,\gamma],\nonumber\eea {\sl i.e.} one recovers
Eq. \eqref{varGutz} obtaining a natural orbitals Gutzwiller
functional.  Since Gutzwiller theory captures many aspects of
correlated systems this suggest that RDMFT may indeed be a promising
root to tread materials in which correlations dominate.


\section{Conclusions}

In this work we have shown how to determine the optimum basis orbitals to
expand a single-band Hubbard-like lattice model starting from a
continuum Hamiltonian. 
The method can be seen as a
multiconfigurational generalization of the Hartree-Fock method where
the solution of the  lattice Hamiltonian provides the configurations
that are included in the wave-function and the OLMO equations provide
the analogy of the Hartree-Fock equations for the orbitals. In the
case in which only one Slater determinant is retained in the lattice
wave-function the OLMO equations reduce to the Hartree-Fock equations for the orbitals as these are
the optimum orbitals for a single Slater determinant. 
In the general case, the ground state can be seen as a variational wave function where both
the orbitals and the weight of the different Slater determinants are
optimized.

As explained above, differently from previous works [\onlinecite{marzari2012,paul2006}] aimed at constructing the
optimum Wannier orbitals for a given truncated lattice Hamiltionian
starting from a fixed  single-particle subspace, here we do not make
any \emph{a priori} assumption on the lattice model and we leave full
variational freedom to the orbitals in order to consider the
optimization of the single-particle subspace that will be used to
construct the lattice model. The procedure requires the simultaneous
determination of the lattice ground-state and the solution of the OLMO
continuum equations. Therefore, the optimum orbitals generally depend on the
lattice ground-state and {\sl viceversa}.

The OLMO equations are quite hard to solve in general thus,  
 it is of great interest to clarify, at least in one test-ground case,
 what the effect of orbital optimization would be. 
 By solving the OLMO equations for a toy-model with two delta-function-interacting fermions in a one dimensional two-well potential, we demonstrate that orbital optimization:
\begin{itemize}
\item[(i)] induces qualitative and quantitative modifications of the parameters of the lattice model;
\item[(ii)] may affect significantly the ground-state of the system
  and excitation energies;
\item[(iii)] in the infinitesimal interaction limit yields orbitals
  and parameters that differ substantially from the non-interacting ones.
\end{itemize}
As regards point (i) we have shown in particular that orbital optimization
tends to increase the ratio $U/t$  while it strongly quenches the
direct exchange coupling $K$ with dramatic implications for the
stability of a ferromagnetic state.

As regards point (ii) we have shown that orbital optimization carried
on ``blindly''  for any state, leads to wrong excitation energies and
a wrong phase diagram since the variational \textsl{Ansatz} has not the same
quality for all
states. In our case it is exact for a triplet state with only contact
interactions among fermions and it is only approximate for singlet states.
Therefore, errors do not cancel when computing excitation energies or
the phase diagram. For Coulomb or other interactions which are large
at short distances we expect a similar state-dependent quality of the variational
state. For example on a single band system of electrons
the exchange hole will reduce much more the Coulomb interaction energy
in the case of the fully polarized state than in the paramagnetic state.
Thus the paramagnetic state is a harder problem for the effective variational
wave-function and the same problem will arise.

We have shown that the best strategy is to choose one reference state
and use the same parameters for all states. The reference state, in our case,
is the singlet state but would be a paramagnetic state in the case of
many atoms. Then the orbitals and parameters in the model are fixed for
all states. This is reassuring for lattice model methods which usually
assume a fixed orbital in the model rather than changing the orbital
(and the Hamiltonian parameters for each state). On the other hand, we
have found that for
some parameters like the direct exchange energy, neglecting orbital
optimization leads to gross errors so the choice of the reference state
is also crucial.

Unfortunately, there is no a general rule to choose what state to use as a
reference state without knowing \textsl{a priori} the ground state of the
system. However, a good guess can be obtained by computing the total
energy in competing states. We see from Fig.~\ref{fig:renpar}c
that there is a modest worsening of the energy by computing the
triplet energy with singlet orbitals and the error depends weakly on the
coupling. On the contrary, computing the singlet energy with the
orbitals optimized for the triplet (NIO) leads to an error (evaluated
as difference with the best singlet variational state) that
diverges at strong coupling. Such an analysis clearly suggests that the
singlet reference state is the best compromise for all states of the
system, as we indeed find, and can be used as a guideline in more
complicated situations.

Another problem in a more general context is how to solve, at least
approximately, the OLMO equations in  many orbital cases.
One possibility is to use some
approximate but accurate lattice solver like dynamical mean-field
theory or a Gutzwiller variational wave-function for the lattice
problem and insert the many-body correlations in the OLMO equations
for the orbitals and iterate until convergence. In this context, it is
interesting that in the case of two-electrons the variational problem
can be mapped to
 a reduced density matrix functional theory which establish a
 connection between this emerging method of continuous approaches and
 the Gutzwiller method, popular in the lattice Hamiltonian community.
  Clearly it  would be worth exploring how far such connection can be
  pushed in extended systems. On the other hand, the link with  
  Gutzwiller theory  puts the RDMFT in a framework which is known to give a good description of the physics of strongly correlated systems.

\section*{Acknowledgements}  
This work was supported by the Italian Institute of
Technology through the project NEWDFESCM and Italian MIUR under
project PRIN-RIDEIRON-2012X3YFZ2. V.B. acknowledges financial support
by Italian MIUR under project FIRB-HybridNanoDev-RBFR1236VV and Premiali-2012 AB-NANOTECH. Z.-J.Y. acknowledges the financial support from the Future and
Emerging Technologies (FET) programme within the Seventh Framework Programme for Research of the European Commission, under FET-Open Grant Number: 618083 (CNTQC).

\bibliographystyle{prsty_no_etal}

\end{document}